\documentclass{PoS}
\usepackage{amsmath}
\usepackage{amssymb}

\usepackage{graphicx}

\title{A 2.5-5 $\mu$m Spectroscopic Study of Hard X-ray Selected AGNs using AKARI InfraRed Camera}

\ShortTitle{AKARI IRC spectra of hard X-ray selectd AGNs}

\author{A.\ Castro$^{a}$, \speaker{T.\ Miyaji}$^{a,b}$, T.\ Nakagawa$^{c}$, M.\ Shirahata$^{c}$, S.\
Oyabu$^{d}$, M.\ Imanishi$^{e}$, Y.\ Ueda$^{f}$, K.\ Ichikawa$^{f}$\\
         \llap{$^{a}$}  Instituto de Astronom\'ia sede Ensenada, UNAM, Ensenada, Mexico\\
         \llap{$^{b}$}  Center for Astrophysics and Space Sciences, UCSD, La Jolla, USA \\
         \llap{$^{c}$}  Institute of Space and Astronautical Science, JAXA, Sagamihara, Japan \\
         \llap{$^{d}$}  Department of Physics, Nagoya University, Nagoya, Japan \\
         \llap{$^{e}$}  Subaru Telescope, NAOJ, Hilo, USA \\
         \llap{$^{f}$}  Department of Astronomy, Kyoto University, Kyoto, Japan\\  
            E-mail: \email{acastro@astrosen.unam.mx}, \email{miyaji@astrosen.unam.mx}
}

\abstract{
We present results of the 2.5-5 $\mu$m  spectroscopy of a sample of hard X-ray selected active galactic nuclei (AGNs) 
using the grism mode of the InfraRed Camera (IRC) on board the infrared astronomical satellite {\sl AKARI}. The sample is selected from the 
9-month Swift/BAT survey in the 14-195 keV band, which provides a fair sample of AGNs including highly absorbed ones. 
 The 2.5-5 $\mu$m spectroscopy provide a strong diagnostic tool for the 
circumnuclear environment of AGNs through the continuum shapes and emission/absorption features such as  the 3.3 $\mu$m 
polycyclic aromatic hydrocarbon (PAH) emission and the broad 3.1 $\mu$m H$_2$O ice, 3.4 $\mu$m  bare carbonaceous dust, 
4.26 $\mu$m CO$_2$ and 4.67 $\mu$m CO absorptions.
 As our first step, we use the 3.3 $\mu$m  PAH emission as a proxy for the star-formation activity and searched for possible difference of star-formation 
activity between type 1 (unabsorbed) and type 2 (absorbed) AGNs. We found no significant dependence of the 3.3 $\mu$m  
PAH line luminosity, normalized by the black hole mass, on optical AGNs types or the X-ray measured column densities.
}

\FullConference{Nuclei of Seyfert galaxies and QSOs - Central engine \& conditions of star formation,\\
		November 6-8, 2012\\
		Max-Planck-Insitut f\"ur Radioastronomie (MPIfR), Bonn,Germany}

\begin{document}

\section{Description of the Research}
\label{sec:description}

Infrared spectroscopy is a powerful tool for the diagnostic buried AGNs, because the effect of dust extinction is relatively small
\cite{imanishi10}. It allows to study sources in which both an AGN and a starburst (SB) are present. Very hard X-ray surveys such as available with {\sl Swift}
Burst Alert Telescope (BAT), provide efficient way of constructing a fair sample of  AGNs including heavily obscured ones. The infrared 2.5-5 $\mu$m 
spectra available with the Infrared Camera (IRC) grism mode on board {\sl AKARI}  provide us with rich pieces of information, which enable us to 
probe the environments of AGNs as well as their circum-nuclear star-formation activities through the continuum shape as well as emission/absorption 
lines (Fig. 1). For example, the 3.3 $\mu$m polycyclic aromatic hydrocarbon (PAH) feature is a good indicator of the star-formation activities and 
the broad 3.1 $\mu$m H$_2$O ice, 3.4 $\mu$m  bare carbonaceous dust, 4.26 $\mu$m CO$_2$ and 4.67 $\mu$m  CO provide information on 
the physical properties of the absorbing material.
\begin{figure}[!ht] \label{fig:spectra}
\begin{center}
\resizebox{\hsize}{!}{
  \centering
  \includegraphics[height=0.35\hsize]{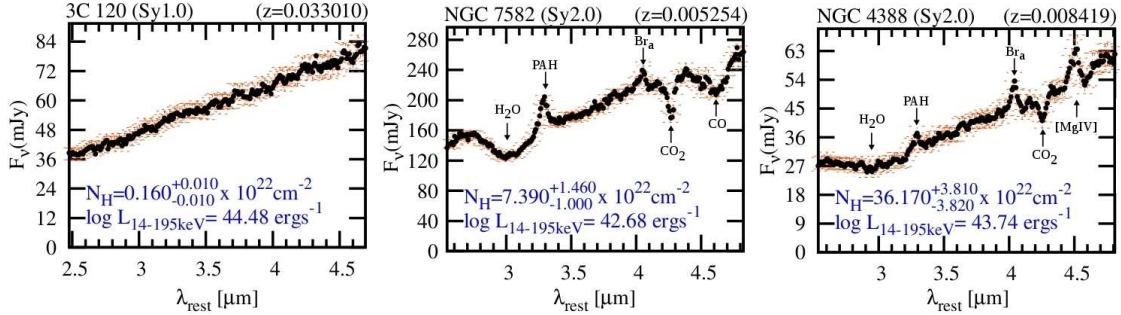}

}\vspace{-\baselineskip}
\end{center}
 \caption{AKARI/IRC infrared 2.5-5 $\mu$m spectra of a random selection of AGNs from our sample. The spectra can reveal the presence starburst, dust obscuration
and AGN free-continuum.  Luminosities  were calculated using $H_{\circ}=70 kms^{-1} Mpc^{-�1}, {\Omega}_m=0.3$ and ${\Omega}_{\Lambda}=0.7$.
   }
\end{figure}

In this work, we investigate the 2.5-5 $\mu$m low-resolution (R $\sim$120) spectra obtained with the IRC instrument of the Japanese space
infrared observatory {\sl AKARI} of a sample of 55 AGNs with various obscuring levels  selected from the 9-month catalog\cite{tueller08} of the
Swift BAT survey, which is sensitive to very high X-ray energies (14-195 keV). Our selected sample also has detailed X-ray spectroscopy (0.3-12
keV)\cite{winter09,ichikawa12}. For all objects in our sample X-ray-derived neutral hydrogen column densities obtained mainly 
by analyzing spectra from {\sl XMM-Newton}, {\sl ASCA}, {\sl Suzaku}, and {\sl Swift}/XRT.  Figure 2 (\emph{Left}) shows the $N_{\rm H}$ histogram 
of the sample.

As our first step, we use the 3.3 $\mu$m PAH emission detected in our spectra (or its upper limits) as a proxy for the star-formation 
activity. We compare the PAH 3.3 $\mu$m luminosities ($L_{3.3 \mu m}$), normalized by the blackhole mass ($M_{\bullet}$) estimates
by a scaling relation with
the J and K band luminosities \cite{mushotzky08},
between Type 1 (un-absorbed) and Type 2 (absorbed) AGNs. 
\begin{figure} \label{fig:distributions} 
\begin{center}
\resizebox{\hsize}{!}{
  \centering
  \includegraphics[height=0.21\hsize]{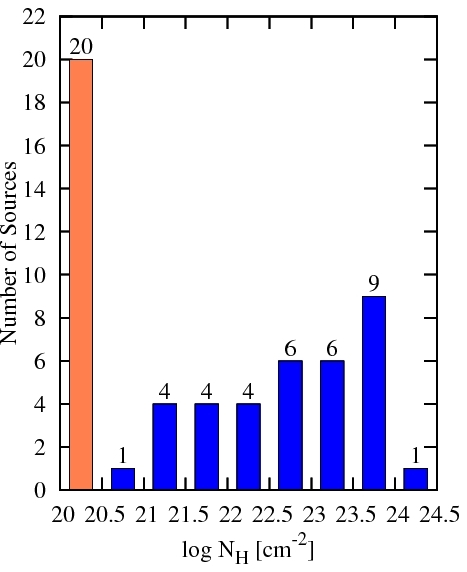}
  \includegraphics[height=0.2\hsize]{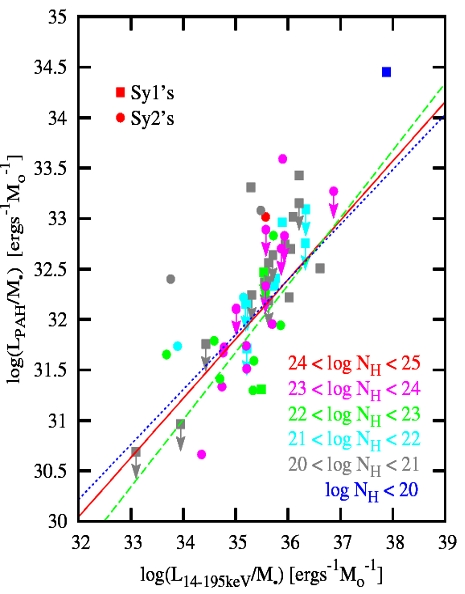}
\includegraphics[height=0.2\hsize]{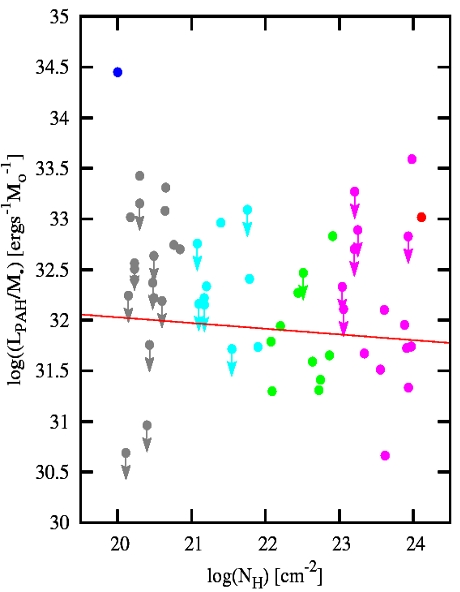}
}\vspace{-\baselineskip}
\end{center}
 \caption{(\emph{Left}) Distribution of the absorbing column density ($N_{\rm H}$) derived from softer X-ray ($E<10$ keV) spectra of our
sample\cite{winter09,ichikawa12}. The shaded histogram at $20<\log N_{\rm H}<21$ bin represents unabsorbed AGNs. (\emph{Center})
$\log(L_{14-�195keV}/M_{\bullet})$ versus $\log(L_{3.3{\mu}m}/M_{\bullet}) $  and  (\emph{right}) $\log N_{H}$ versus $\log(L_{3.3{\mu}m}/M_{\bullet})$
relations. The best-fit lines from linear regressions with upper limits are also shown. 
  } 
\end{figure}

\section{Results}
To test the difference of star-formation activities between different  types  of  AGNs  the  following correlation  analyses  
have  been  made.  The  regression  has  been  made  using  the EM algorithm (included in the ASURV package\cite{isobe86}) to 
account for upper-limits of the $L_{3.3{\mu}m}$.

\begin{enumerate}
 \item Compare  the  log($L_{3.3{\mu}m}/M_{\bullet}$)-log($L_{14-�195keV}/M_{\bullet}$) correlations  of  optical  Type  1  and  Type  2  AGNs
(Fig. 2 [\emph{Center}]). Luminosities are in units of $[{\rm erg\,s^{-1}}]$  and $M_\bullet$ in units of $[{\rm M_{\rm \odot}}]$. 

\begin{center}
Type 1 (28 objects): $\log(L_{3.3{\mu}m}/M_{\bullet}) \sim  (0.67\pm0.19)\log(L_{14-�195keV}/M_{\bullet})+( 8.340\pm6.73)$ \\
Type 2 (27 objects): $\log(L_{3.3{\mu}m}/M_{\bullet}) \sim  (0.54\pm0.17)\log(L_{14-�195keV}/M_{\bullet})+( 12.78\pm6.02)$
\end{center}

\item Investigate the relation between $\log(L_{3.3{\mu}m}/M_{\bullet})$ and $\log N_H$ (Fig. 2. [\emph{Right}]). 
The result is:
\begin{center}
$\log(L_{3.3{\mu}m}/M_{\bullet}) \sim  (-0.05\pm0.08)\log(N_{H})+( 33.15\pm1.89)$ 
\end{center}

\end{enumerate}
\section{Conclusions}
We investigate the 2.5-5 $\mu$m spectra of 55 bright nearby AGNs from the 9-month {\sl Swift} BAT catalog using {\sl AKARI}/IRC. We 
investigate the relation between AGN type/absorption and star formation activities. From our analysis,    
we have found no significant difference between the star formation and the optical AGN type. We have found no significant 
correlation between the neutral gas absorption towards the AGN measured by X-ray absorption and circum-nuclear 
starburst activity.

\end{document}